\documentclass[12pt]{article}
\usepackage{amsmath}
\usepackage{amssymb}

\begin{document}
\begin{flushright}
KFKI-1988-05/A\\
HU ISSN 0368 5330\\
February 1988
\end{flushright}
\vskip 3truecm
\begin{center}
CONTINUOUS QUANTUM MEASUREMENT\\ AND IT\^O FORMALISM
\vskip 1truecm
L. Di\'osi
\vskip .5truecm
Central Research Institute for Physics\\
H-1525 Budapest 114, P.O.B. 49, Hungary
\end{center}
\vskip 1truecm
ABSTRACT

A new quantum-stochastic differential calculus is derived for representing 
continuous quantum measurement of the position operator.
Closed nonlinear quantum-stochastic differential equation is given
for the quantum state of the observed particle. 
Possible physical realization of continuous position measurement
is constructed.
\pagenumbering{gobble}
\newpage
\pagenumbering{arabic}
{\bf 1. Introduction}
\vskip 2cm
In 1982 Barchielli, Lanz and Prosperi suggested  a theory
of continuous quantum measurement$^1$ including observables of
continuous spectra such like, e.g., coordinate operator.
Various equivalent formalisms were elaborated: path 
integrals$^1$, characteristic functionals$^2$, quantum-stochastic
differential equations (QSDE's)$^{3,4}$. Very recently, Ghirardi,
Rimini and Weber have proposed the appealing unification of
microscopic and macroscopic dynamics$^5$ in the frame of a
similar theory to that of continuous quantum measurements.

In this paper we suggest a simple quantum-stochastic 
formulation, \`a la It\^o, for continuous quantum measurements$^1$.
It shows very close relationship to the Stratonowitch-calculus 
used by Gisin for discrete observables$^6$. In Sect.
2 we outline the principles of continuous observation and
present the proper nonlinear It\^o-equations (which differ 
from It\^o-equations of Ref. 4). The derivation  of our
equations is postponed to Sect. 3; in Sect. 4 we propose a
possible physical model to which our formalism can 
presumably be applied. 
\newpage
{\bf 2. Continuous position measurement, It\^o-equations}
\vskip 2cm
In order to reduce the amount of technical work we single out
the simplest example and we consider the \emph{continuous
measurement} of the position $\hat q$ of a free particle of mass $m$
moving in one dimension.

Let the state vector $\psi$ satisfy the free Schr\"odinger
equation; then the density operator $\hat\rho\equiv\psi\psi^\dagger$ obeys
the following equation:
$$
\hat\rho~=~-(i/2m)[\hat p^2,\hat\rho]~~~;
\eqno(2.1)
$$
$\hat p$ is the momentum operator canonically conjugated to the
coordinate $\hat q$. Now, Refs. 1, 5 introduce instantaneous
stochastic changes, too:
$$
\hat\rho~\rightarrow~[P(\bar{q})]^{-1}T_{\bar {q}}[\hat\rho]~\equiv~~~~~~~~~~~~~~~~~~~~~~~~~~~~~~~~~~~~~~
$$
$$
\equiv~[P(\bar{q})]^{-1}\sqrt{\alpha/\pi}\exp[-{\scriptstyle{1\over2}}\alpha(\hat q-\bar q)^2]
\hat\rho \exp[-{\scriptstyle{1\over2}}\alpha(\hat q-\bar q)^2]
\eqno(2.2)
$$
called \emph{measurement} (or \emph{localization}) processes which
repeatedly occur at equidistant moments$^1$ separated by
$\Delta t$. In the above equation $\alpha$ denotes the accuracy
parameter of measurement. The \emph{measured coordinate} $\bar q$ is
selected at random with the probability distribution
$$
P(\bar{q})~=~ \mathrm{tr}T_{\bar {q}}[\hat\rho]~=~
\sqrt{\alpha/\pi}\mathrm{tr}\hat\rho\exp[-\alpha(\hat q-\bar q)^2]~.
\eqno(2.3)
$$
\newpage
\noindent The Eqs. (2.1-3) prescribe a certain joint stochastic
process for $\hat\rho\equiv\psi\psi^\dagger$ and $\bar q$. Barchielli et al.
define the \emph{continuous measurement} of the position $\hat q$ by
taking the following limit
$$
\Delta t,\alpha~\rightarrow~0,~\alpha/\Delta t~=\gamma~=\mathrm{const}.
\eqno(2.4)
$$
of the stochastic process. [Here we have to note that
Ghirardi et al.$^5$ suggest \emph{finite} values for $\alpha$ and $\Delta t$,
without requiring the limes (2.4). Nevertheless, they also
notice that most physical characteristics of their theory
depend only on the ration $\gamma=\alpha/\Delta t$.]

Observe that, while $\hat\rho$ is a function of the time $t$,
the other stochastic variable $\bar q$ only makes sense for the
instants of the measurement process (2.2). Let us extend
the definition of $\bar q$ for all time: one can, e.g., identify $\bar q_t$
by the proper continuous zig-zag function of time. We
introduce then a new variable $\bar Q$ instead of $\bar q$:
$$
\bar Q_t~=\int\limits_0^t \bar q_{t'}dt'~~.
\eqno(2.5)
$$

In the next Section we shall prove that $\hat\rho$ \emph{and} $\bar Q$
\emph{follow a Gaussian process}. They obey the couple of
\emph{stochastic differential equations}; one is for the $\hat\rho$-valued
quantum-stochastic process:
\newpage
$$
d\hat\rho~=~(-i[\hat p^2/2m,\hat\rho]
-{\scriptstyle{1\over4}}\gamma[\hat q,[\hat q,\hat\rho]])dt~+~\{\hat q-\bar q,\hat\rho\}d\xi~,
\eqno(2.6)
$$
the other is for the \emph{measured coordinate}:
$$
d\bar Q~=~\langle\hat q\rangle dt~+~ \gamma^{-1}d\xi
\eqno(2.7)
$$
Here $\xi$ is a Wiener-process. The \emph{It\^o-differential}\/$^7$ $d\xi$
satisfies the following algebra:
$$
\langle d\xi \rangle_\mathrm{st}~=~0~,
$$
$$
d\xi d\xi~=~{\scriptstyle{1\over2}}\gamma dt~,
\eqno(2.8)
$$
$$
(d\xi)^n~=~0~~\mathrm{if}~n=3,4,\dots
$$
Through our paper $\langle .\rangle$ stands for the quantum expectation 
values while $\langle .\rangle_\mathrm{st}$ denotes stochastic means.

It is important to see that the QSDE (2.6) preserves the 
pure state property $\hat\rho\equiv\psi\psi^\dagger=\hat\rho^2$. In fact, it is
enough to prove that $d\hat\rho^2=d\hat\rho$ provided $\hat\rho^2=\hat\rho$ at
a given moment. We therefore substitute the QSDE (2.6) into
the RHS of the identity $d\hat\rho^2=\{d\hat\rho,\hat\rho\}+d\hat\rho d\hat\rho$.
Using the It\^o-algebra (2.8) and the assumption  $\hat\rho^2=\hat\rho$,
we arrive at the identity $d\hat\rho^2=d\hat\rho$.

One may notice that the so-called measured coordinate $\bar q$
is not a good representative of the particle trajectory.
From the Eqs. (2.5) and (2.7) one can formally get
\newpage
$$
\bar q~=~\langle\hat q\rangle~+~\gamma^{-1}~\dot\xi
\eqno(2.9)
$$
which means that $\bar q$ is charged by a stationary white-noise
making the graph $\bar q_t$ an awkward fractal$^8$ instead of a
trajectory. This problem has already been emphasized in
Refs. 1, 5 as well.

It is thus better to represent the trajectory by $\langle\hat q\rangle$
instead of $\bar q$. Consequently, our proposal is to concentrate
on the \emph{nonlinear quantum-stochastic differential equation 
(QSDE)} (2.6) [together with the It\^o-algebra (2.8)]. This
QSDE, in itself, will account for all the physical behaviour
of the particle affected by the continuous coordinate
measurement.
\vskip 2cm
{\bf 3. Verification of It\^o-equations}
\vskip 2cm

In the present Section we are going to derive the
stochastic differential equations (2.6),(2.7) starting from
Eqs. (2.1-4) of continuous position measurement$^1$.

First we need the notion of stochastic mean for the
measurement process (2.2):
$$
\langle~.~\rangle_\mathrm{st}~\equiv~\int .~ P(\bar q)d\bar q
\eqno(3.1)
$$ 

We introduce the change of our stochastic variables $\hat\rho$
and $\bar Q$ taken for one cycle of duration $\Delta t$:
\newpage
$$
\Delta\hat\rho_t~\equiv~\hat\rho_{t+\Delta t}-\hat\rho_t~=~~~~~~~~~~~~~~~~~~~~~~~~~~~~~~~~~~~~~~~~~~~~~~~~~~~~~~~~~~~~~~~
$$
\vskip -42pt
$$
=~-i[\hat p^2/2m,\hat\rho_t]\Delta t~+~[P_t(\bar q_t)]^{-1}T_{\bar q_t}[\hat\rho_t]
                                                                          -\hat\rho_t~+~\mathcal{O}[(\Delta t)^2]~~~~~~~~~~~
\eqno(3.2)                                                                          
$$
$$
\Delta\bar Q_t~\equiv~\bar Q_{t+\Delta t}-\bar Q_t~=~\bar q_t\Delta t~ +~\mathcal{O}[(\Delta t)^2]~~~~~~~~~~~~~~~~~~~~~~~~~~~~~~
\eqno(3.3)
$$
where we used Eqs. (2.1-2) and Eq. (2.5), respectively.

The continuous measurement process (2.1-4) is a Gaussian
process with It\^o-equations (2.6-8) if, in the limes (2.4),
the moments of $\Delta\hat\rho$ and $\Delta\bar Q$ [c.f. Eqs. (3.2),(3.3)]
satisfy the same algebra as the corresponding It\^o-differentials 
$d\hat\rho$, $d\bar Q$ [c.f. Eqs. (2.6),(2.7)].
Consequently, we have to prove the following asymptotics:
$$
(1/\Delta t)\langle\Delta\hat\rho\rangle_\mathrm{st}\rightarrow
(1/dt)\langle d\hat\rho\rangle_\mathrm{st}
~=~-i[\hat p^2/2m,\hat\rho]-{\scriptstyle{1\over4}}\gamma[\hat q,[\hat q,\hat\rho]]~~~~~~~~~~~~~
\eqno(3.4a)
$$
$$
(1/\Delta t)\langle\Delta\bar Q\rangle_\mathrm{st}\rightarrow
(1/dt) d\bar Q
~=~ \langle\hat q\rangle~~~~~~~~~~~~~~~~~~~~~~~~~~~~~~~~~~~~~~~~~~~~
\eqno(3.4b)
$$
$$
(1/\Delta t)\langle\Delta\hat\rho\otimes\Delta\hat\rho\rangle_\mathrm{st}\rightarrow
(1/dt) d\hat\rho\otimes d\hat\rho
~=~{\scriptstyle{1\over2}}\gamma\{\hat q-\bar q,\hat\rho\}\otimes\{\hat q-\bar q,\hat\rho\}~~
\eqno(3.4c)
$$
$$
(1/\Delta t)\langle\Delta\bar Q\Delta\bar Q\rangle_\mathrm{st}\rightarrow
(1/dt) d\bar Q d\bar Q
~=~1/2\gamma~~~~~~~~~~~~~~~~~~~~~~~~~~~~~~~~
\eqno(3.4d)
$$
$$
(1/\Delta t)\langle\Delta\bar Q\Delta\hat\rho\rangle_\mathrm{st}\rightarrow
(1/dt) d\bar Q d\hat\rho
~=~{\scriptstyle{1\over2}}\{\hat q-\bar q,\hat\rho\}~~~~~~~~~~~~~~~~~~~~~~~~~~
\eqno(3.4e)
$$
$$
(1/\Delta t)\langle\mbox{higher than 2nd powers of $d\hat\rho$ and/or $d\bar Q$}\rangle_\mathrm{st}\rightarrow~0~.~~~~~~~~~~~~~
\eqno(3.4f)
$$
The stochastic means on the LHS's must be taken at time $t$
(i.e. at the beginning of the interval $\Delta t$). The
expressions on the very right have been calculated by using
the stochastic differential equations (2.6),(2.7) and the
It\^o-algebra (2.8). To prove the above asymptotical 
relations, we are going to evaluate the LHS terms in turn.
\newpage

As for the first one, Eq. (3.4a) is known from previous
works. Its proof is easy because the probability
distribution $P(\bar q)$ cancels from $\langle\Delta\hat\rho\rangle_\mathrm{st}$. Thus, we
refer the reader to the literature$^{1,5}$ where the following
master equation has been proved:
$$
d/dt\langle\Delta\hat\rho\rangle_\mathrm{st}
~=~-i[\hat p^2/2m,\langle\Delta\hat\rho\rangle_\mathrm{st}]
-{\scriptstyle{1\over4}}\gamma[\hat q,[\hat q,\langle\Delta\hat\rho\rangle_\mathrm{st}]]~.
\eqno(3.5)
$$
Its special case, i.e. when 
$\langle\Delta\hat\rho\rangle_\mathrm{st}=\psi\psi^\dagger=\hat\rho$ at
time $t$, yields just the Eq. (3.4a).

To prove Eqs. (3.4b-f) we need the form of the
distribution $P(\bar q)$. In what follows we exploit translational
invariance and choose a special coordinate system where
$\langle\hat q\rangle=0$ for the give  state $\hat\rho$ at time $t$; this choice makes
our calculations simpler without loss of their generality.
Then Eq. (2.3) yields the following asymptotical expansion:
$$
P(\bar q)~=~\sqrt{\alpha/\pi}\exp(-\alpha\bar q^2)
[1+\mathcal{O}(\alpha)+\mathcal{O}(\alpha^2\bar q^2)]~.
\eqno(3.6)
$$
In this approximation, Eq. (3.1) yields
$$
\langle\bar q\rangle_\mathrm{st}~=~\mathcal{O}(\alpha)~~~~~~~~~~~~~~~~
\eqno(3.7a)
$$
$$
\langle\bar q^2\rangle_\mathrm{st}~=~(1/2\alpha)~+~\mathcal{O}(1)~.
\eqno(3.7b)
$$
and, by virtue of Eq. (3.2), we obtain
$$
\Delta\hat\rho~=~\alpha\bar q\{\hat q,\hat\rho\}~+~\mathcal{O}(\alpha^2\bar q^2)
\eqno(3.8)
$$
\newpage
\noindent We remind the reader that we have supposed $\langle\hat q\rangle=0$.

Let us evaluate each LHS term of Eqs. (3.4b-e) in turn,
with the help of Eqs. (3.3), (3.7ab) and (3.8):
$$
(1/\Delta t)\langle\Delta\bar Q\rangle_\mathrm{st}\rightarrow
(1/\Delta t)\langle\bar q\rangle_\mathrm{st}\Delta t~\rightarrow~ 0~~~~~~~~~~~~~~~~~~~~~~~~~~~~~~~~~~~~~~~~~~~~~~~~~~
$$
$$
(1/\Delta t)\langle\Delta\hat\rho\otimes\Delta\hat\rho\rangle_\mathrm{st}\rightarrow
(1/\Delta t) \alpha^2\langle\bar q^2\rangle_\mathrm{st}\{\hat q,\hat\rho\}\otimes\{\hat q,\hat\rho\}\rightarrow{\scriptstyle{1\over2}}\gamma\{\hat q,\hat\rho\}\otimes\{\hat q,\hat\rho\}~~~~~
$$
$$
(1/\Delta t)\langle\Delta\bar Q\Delta\bar Q\rangle_\mathrm{st}\rightarrow
(1/\Delta t)\langle \bar q^2\rangle_\mathrm{st}(\Delta t)^2\rightarrow1/2\gamma~~~~~~~~~~~~~~~~~~~~~~~~~~~~~~~~~~~~~
$$
$$
(1/\Delta t)\langle\Delta\bar Q\Delta\hat\rho\rangle_\mathrm{st}\rightarrow
(1/\Delta t) \alpha\Delta t \langle \bar q^2\rangle_\mathrm{st}\{\hat q,\hat\rho\}
\rightarrow{\scriptstyle{1\over2}}\{\hat q,\hat\rho\}~.~~~~~~~~~~~~~~~~~~~~~~~~~~~~~~
\eqno(3.9)
$$
Recalling that $\langle\hat q\rangle=0$, the Eqs. (3.9) have provided the proof
of Eqs. (3.4b-e). We shall not give a systematic proof to
Eq. (3.4f). The reader may convince himself that the
increasing powers of $\Delta t$ (or, equivalently, of $\alpha$) will
not further be compensated by the higher moments of $\bar q$,
and expressions like $1/\Delta t\langle(\Delta\bar Q)^3\rangle_\mathrm{st}$, 
$1/\Delta t\langle(\Delta\hat\rho\otimes\Delta\hat\rho\otimes\Delta\hat\rho\rangle_\mathrm{st}$,
e.t.c. will tend to the zero. Besides, this is the
genuine feature if Gaussian processes.

So we have presented the proof of It\^o-equations
(2.6),(2.7) for the continuous measurement (2.1-4)
of a free particle.
\newpage
{\bf 4. Possible physical model}
\vskip 2cm

We try to construct a physical model whose state would
follow the quantum stochastic process described by the QSDE
(2.6). Let us suppose that our test particle is not free but
immersed in a parallel stream of light particles (e. g. of
photons). We consider the quantum motion of the probe only
in the transversal plane. This model has been stimulated by
a similar discussion given recently by Joos and Zeh$^9$.
(Meanwhile, a more formal treatment has been shown by Caves
and Milburn$^{10}$). 

At the beginning, let us start the probe with a given
pure quantum state $\hat\rho=\psi\psi^\dagger$. It obeys to the free
Schr\"odinger-equation (2.1) until the first collision with a
photon occurs. The  quantum state of the probe bears an
instantaneous random change. It is very crucial to realize
that this change \emph{(collapse) may be identified if we detect the
scattered photon}. The set of possible collapses depends
on the photon measuring apparatus. If, e.g., we observe the
scattered photon through an optical lens, we can detect the
current position of the probe. The non-unitary localization
process (2.2) may, at least qualitatively, represent the
collapse of the quantum state where $1/\sqrt{\alpha}$ is the resolution
of the optical device in the transverse plane. The operator
$\hat q$ stands for the transverse coordinate: the probability
distribution of the measured transverse position $\bar q$ of the
probe is given by (2.3).
\newpage
 
The optimal resolution $1/\sqrt{\alpha}$ and the repetition 
frequency $1/\Delta t$ of single measurements depend on the wave
number $k$ of photons, on the intensity $I$ of their stream and
on the total cross section $\sigma$ of photon scattering on the
probe:
$$
1/\sqrt{\alpha}~\approx~1/k,~~~~~~~1/\Delta t~\approx~\sigma I~.
\eqno(4.1)
$$

Now, provided the current wave function of the probe is
such that 1) its transverse width is much smaller than the
resolution $1/\sqrt{\alpha}$, 2) its relative change during $\Delta t$ in
between collisions is small and 3) we consider the probe
properties on time scales larger than $\Delta t$ then, we can
exploit the property of the limes (2.4). Hence, one may
expect that the QSDE (2.6) will account for the change of
the probe's quantum state $\hat\rho$ and, furthermore, the
observed position $\bar q$ is governed by the stochastic
differential equation (2.7).

It would be interesting to see experimentally the fractal
nature of the observed ``trajectory''. We should remember,
that this fractality as well as the validity of the Eqs.
(2.6-7) break down at time scales equal or less than $\Delta t$.

Finally, we briefly discuss the alternative continuous
measurement on the very same system. Let us remove the lens
from our measuring apparatus and observe thus the momentum
of the scattered photon. The the quantum state acquires the
\emph{unitary} change
\newpage
$$
\hat\rho~\rightarrow~\exp(i\Delta\bar p\hat q)\hat\rho\exp(-i\Delta\bar p\hat q)
\eqno(4.2)
$$
instead of the collapse (2.2): here $\Delta \bar p$ is the measured
(transverse) momentum transfer whose probability
distribution depends on the differential cross section of
the collision. We note without proof that, in the limes
(2.4), the following QSDE fulfils for the quantum state of
the probe:
$$
d\hat\rho~=~(-i[\hat p^2/2m,\hat\rho]-{\scriptstyle{1\over4}}\gamma[\hat q,[\hat q,\hat\rho]])dt~-i[\hat q,\hat\rho]d\xi~.
\eqno(4.3)
$$
This QSDE is \emph{linear} and it corresponds to \emph{unitary} evolution
of the probe state in the effective white-noise potential
$\hat q\dot\xi$ of the photon ``heat bath'', c.f. Refs. 11, 12.

Note that both QSDE's (2.6) and (4.3) yield the same
master equation (3.5) since the observation of scattered
photon does not make any change to the statistical operator
$\langle\hat\rho\rangle_\mathrm{st}$ of the probe. The general pure state QSDE (i.e.
preserving $\hat\rho\equiv\hat\rho^2$) for the same system has the 
following stochastic term:
$$
\cos\beta\{\hat q-\langle\hat q\rangle,\hat\rho\}d\xi~-~i\sin\beta[\hat q,\hat\rho]d\xi
\eqno(4.4)
$$
with arbitrary real $\beta$. In particular, this would correspond
to the simultaneous unsharp measurement of the position and
the momentum transfer of the \linebreak
\newpage
\noindent probe. (We mention the \emph{non-Gaussian} pure state stochastic process$^{13}$ which could
probably be connected with some even more sophisticated
measurement of the scattered photon.)
\vskip 2cm
{\bf 5. Concluding remarks}
\vskip 2cm

The formal extension of the It\^o-equations (2.6-8) for the
continuous measurement of several (not necessarily
commuting) operators seems to be straightforward. It would,
nevertheless, be desirable to recapitulate the proof 
starting, e.g., from the mathematical representation of
continuous measurements presented in Refs. 3, 4.

In a forthcoming paper we shall analyze the evolution of
the QSDE (2.6) and we shall show that it possess a unique
stationarily localized solution suitable to represent
classical trajectories. 
\vskip 2cm
The author thanks Profs. A. Frenkel and B. Luk\'acs for helpful discussions
and Prof. P. Hrask\'o for useful remarks.
\newpage

\parskip 0truecm
\vskip 0.5 truecm
\noindent
{\bf References}
\vskip 2cm

\noindent\hskip 20pt [1] A.Barchielli, L.Lanz and G.M.Prosperi, Nuovo Cim. {\bf 72B}, 79 (1982)
\vskip .2 truecm

\noindent\hskip 20pt [2] A.Barchielli, Nuovo Cim. {\bf 74B}, 113 (1983)
\vskip .2 truecm

\noindent\hskip 20pt [3] A.Barchielli and G.Lupieri,  J.Math.Phys. {\bf 26}, 2222 (1985)
\vskip .2 truecm

\noindent\hskip 20pt [4] A.Barchielli, Phys.Rev. {\bf D34}, 1642 (1986)
\vskip .2 truecm

\noindent\hskip 20pt [5] G.C.Ghirardi, A.Rimini and T.Weber, Phys.Rev. {\bf D34}, 470 (1986)
\vskip .2 truecm

\noindent\hskip 20pt [6]  N.Gisin, Phys.Rev.Lett. {\bf 82}, 1657 (1984) 
\vskip .2 truecm

\noindent\hskip 20pt [7] L.Arnold, Stochastic Differential Equations: Theory
and Applications 

\noindent\hskip 35pt (Wiley, New York, 1974)
\vskip .2 truecm

\noindent\hskip 20pt [8] L.F.Abbot and M.B.Wise, Am.J.Phys. {\bf 49}, 37 (1981)
\vskip .2 truecm

\noindent\hskip 20pt [9] E.Joos and H.D.Zeh,  Z.Phys. {\bf B59}, 223 (1985)
\vskip .2 truecm

\noindent\hskip 20pt [10] C.M.Caves and G.J.Milburn, Phys.Rev. {\bf A36}, 5543 (1987)
\vskip .2 truecm

\noindent\hskip 20pt [11] V. Gorini et al., Rep. Math. Phys.  {\bf 13} 149 (1987)
\vskip .2 truecm

\noindent\hskip 20pt [12] V.R.Chechetkin and V.S.Lutovinov, J.Phys. {\bf A20}, 4757 (1987)
\vskip .2 truecm

\noindent\hskip 20pt [12] L.Di\'osi, Phys.Lett. {\bf 112A}, 288 (1985)
\vskip .2 truecm

\end{document}